\documentclass[12pt]{article}
\usepackage{latexsym}
\usepackage{apalike}
\input epsf.tex
\def\simlt{\mathrel{\lower .3ex \rlap{$\sim$}\raise .5ex \hbox{$<$}}}
\def\simgt{\mathrel{\lower .3ex \rlap{$\sim$}\raise .5ex \hbox{$>$}}}
\def\half{ \frac {1}{2} }
\def\xvec{ \vec{x} }

\def\yvec{ \vec{y} }
\def\zvec{ \vec{z} }
\def\jvec{ \vec{j} }
\def\kvec{ \vec{k} }

\def\pvec{ \vec{p} }
\def\tprime{ t^\prime }

\newcommand{\rb}[1]{\raisebox{1.5ex}[0pt]{#1}}

\begin{document}

\title{Analysis of a Population Genetics Model With
Mutation, Selection, and Pleiotropy%\\
%{\small Running title: A pleiotropic population genetics model}
}
\author{
S.N. Coppersmith,\thanks{
The James Franck Institute, The University of Chicago,
5640 S. Ellis Avenue, Chicago, IL  60637}
\hspace{0in}\thanks{
Corresponding author.  Email address: snc@control.uchicago.edu;
Fax number: (773) 702-5863.}
~~Robert D. Blank,\thanks{The Hospital for Special Surgery,
535 East 70th Street, New York, NY  10021}
~~and Leo P. Kadanoff\footnotemark[1]   %%LPK
}
\maketitle
%\vspace{1cm}

\baselineskip = 16pt

\begin{abstract}
\baselineskip = 16pt
We investigate the behavior of a population genetics model
introduced by \cite{waxman98} incorporating
mutation, selection, and pleiotropy.
The population is infinite and continuous variation of genotype is allowed.
Nonetheless,
Waxman and Peck showed that if the degree of pleiotropy
is large enough,
in this model a nonzero fraction of the population can have
identical alleles. This `condensed mode' behavior appears in the limit
of infinite times.

This paper explores the time-dependence of the distribution
of alleles in this model.
First, the model is analyzed using a recursion technique
which enables the distribution of alleles to
be calculated at finite times as well as in Waxman and Peck's
infinite-time limit.
Second, both Waxman and Peck's original model and a related model
in which mutations occur continuously are mapped onto problems
in quantum mechanics.
In both cases,
the long-time analysis for the biological model is equivalent to finding
the nature of the eigenstates
of the quantum problem.  The condensed mode appears if and only if there is
no bound state in the
quantum problem.

We compare the behavior of the discrete- and continuous-time
versions of the model.
The results for the two cases are qualitatively similar,
though there are some quantitative differences.
We also discuss our attempts to
correlate the statistics of DNA sequence variations with the
degree of pleiotropy of various genes.

\end{abstract}
\vskip .5 cm
%{\bf Key words:}  Population genetics~---~Pleiotropy~---~Quantum mechanics 
\vskip .75 cm
\section{Introduction}
\label{introduction}
Recently, \cite{waxman98} introduced a
simple population genetics model of an infinite
population with a continuous distributions of alleles\footnote{%
Waxman and Peck's model is based on the
continuum-of-alleles model of \protect{\cite{kimura65}}.
Other important investigations of pleiotropic models include
\cite{wagner89}, \cite{lande80}, \cite{gavrilets93}, and
\cite{turelli85}.}
incorporating pleiotropy
(one gene affecting several characters of an
organism).
They demonstrated that when the number of characters
affected is greater than two, the long-time steady state solution
of their model can have a nonzero fraction of the population with
identical alleles, and that this phenomenon does not occur if
the number of characters affected per gene is two or fewer.

In this paper we consider their model and a
simple variant of it, addressing several issues.
The first three  %%LPK chANGED
issues involve the
mathematical analysis of the theory.
First we study the time evolution of the distribution
of alleles.
In the model, the initial
distribution of alleles is continuous,
and at infinite time there is an
infinitely narrow ($\delta$-function) peak;\footnote{%
The Dirac delta-function, $\delta(t)$, is defined to be a
function very sharply peaked
around $t=0$, but having a total weight $\int dt \delta(t)$ equal to one.
One definition of such a
function is the limit as $z$ goes to positive infinity of ${(2 \pi
z)}^{-1/2}\exp[-t^2/(2z)]$.
We shall deal with variables $\xvec$ with $N$ components.
For such variables, $\delta(\xvec)$ is
defined to be a delta function in the first component of $\xvec$, times one
in the second component,
$\ldots$.
In this way, one has $\int d^Nx~\delta(\xvec)=1$.\cite{spanier87}}
we ask how the distribution of
alleles depends on time at long but finite times.
%%LPK added
Second, we describe how to construct time-dependent solutions by using an
expansion
in a sum of functions (Gaussians) particularly picked to meet the
requirements of this
problem.
%%LPK added
Third, we examine
the relationship between the model
with discrete generations used by Waxman and Peck (where
mutations occur at discrete times, and fitness selection
occurs continuously) and a continuous-time model
in which
both mutations and fitness selection occur continuously.
The latter is a limiting case of the discrete model arising when the
time between generations gets very short.
We demonstrate that the qualitative behavior of the
two models is the same, but that dependence of the
behavior on the parameters can be different in the two
models.

We show that the
%equations describing the
population genetics models can be mapped onto problems in
quantum mechanics, specifically Bose-Einstein
condensation\footnote{Bose-Einstein condensation in three dimensions
is discussed in many texts on statistical mechanics,
e.g.,~\protect{\cite{feynman72}}.}\cite{degroot50}
and motion of a particle in a central potential.\footnote{
See any book on quantum mechanics, e.g., \protect{\cite{schiff68}}.}
The mapping onto Bose condensation is performed on the
discrete-time model in the limit that the fitness
selection is very strong
so that
only organisms with fitnesses
very near optimum survive each generation.
The mapping onto the quantum particle in a central potential
applies more generally. The simplest case
has the potential independent of time. This case corresponds to the
a limit of the population model
in which fitness selection and mutations both occur continuously.
The close relation between the continuous-time
population biology models and Schr\"odinger's equation has
been exploited by B\"urger\footnote{A recent paper with references
to previous work is \protect{\cite{burger96}}.}
to obtain many general results on
the long-time behavior of models of the type studied here.
In this paper, we focus on the mapping for a specific model,
which yields a simple physical interpretation of the
emergence of a unique genotype, and allows the extraction
of the time-dependence of the fitness peak as well
as comparison to the discrete-time model.
We find that the behavior is qualitatively identical
when the mutations are continuous and when they are
discrete, though there are some quantitative differences.

We also discuss our attempts to relate the results from
this model to the statistical properties of the sequences
that are archived in various sequence databases.
We analyze the statistical properties of gene sequences
tabulated in genetic databases and attempt to correlate
observed variations with estimates of the degree of pleiotropy
of different genes in the databases.
The results of these attempts are inconclusive.

This paper is organized as follows.  In section \ref{models}
we introduce the theoretical models.
Section \ref{discrete} presents our analysis of the time evolution
of the discrete version of the model, demonstrating that the behavior
can be extracted analytically in a limit in which the typical jump caused
by mutations is very large.
In section \ref{continuous} we
present our analysis of the continuous-time model.
Section \ref{crossover} contains remarks on the relationship
between the continuous- and discrete-time models.
Section \ref{database} presents our examination of
DNA sequence archives and our attempts to correlate
the statistics of documented sequence variations
to the degree of pleiotropy of various genes.
Section \ref{conclusion} recaps our main conclusions.
The Appendix contains mathematical details of the analysis of the
time-dependence of the continuous-time model aimed at establishing the
long-time time-dependence.

\section{The Models}
\label{models}
One model we study is that of \cite{waxman98}.
In this model,
the population is infinite, the phenotypic
variation is continuous,
each gene affects $N$ characters, and
the effects of different genes are uncorrelated (no
linkage disequilibrium or epistasis).
The model assumes a very large population of haploid and
asexual organisms with discrete generations.
Let $\xvec$ be a continuous vector with $N$ components
which represents characters which determine the
viability of an organism, and $\phi(\xvec, t)$ be
the normalized probability density that an organism
in the population has the characters given by $\xvec$
at time $t$.
%% LPK added
The normalization condition for the probability is given by an
$N$-dimensional integral as
\begin{equation}
\int \phi(\xvec,t)  d^Nx~ =1~.
\label{norm}
\end{equation}
%% LPK added
At each generation
an organism with characters $\xvec$ survives
viability selection with a probability proportional to
$\exp(-x^2/2V_s)$.
%% LPK added
This Gaussian fitness selection will play
an essential role in our solution of the model.
%%LPK added
Once each generation, a fraction $\theta$ of the
population mutates;
it is assumed that if a mutation
occurs, then the probability that the mutant takes on
the value $\xvec$ given the parental value $\yvec$
is $f(\xvec - \yvec)$.

Since $\phi(\xvec, t)$ is a probability density, it is
normalized at every time $t$.
Thus, for this model,
the equation for the time evolution of $\phi(\xvec, t)$ is:
\begin{equation}
\Gamma(t+1) \phi(\xvec, t+1) = (1 - \theta) w_1(\xvec)\phi(\xvec, t)  %%
%LPK changed
+ \theta \int f(\xvec - \yvec) w_1(\yvec) \phi(\yvec,t) d^Ny ~,
\label{wpequation}
\end{equation}
where
the fitness factor $w_1(\xvec)$ is  %%LPK ADDED
\begin{equation}
w_1(\xvec) = \exp[-x^2/2V_s]~.
\label{w1def}
\end{equation}
%% LPK added
The multiplicative factor, $\Gamma(t+1)$, in Eq.~(\ref{wpequation})
ensures that the probability,
$\phi(\xvec, t+1)$ is properly normalized at every time step.
Integrating Eq.~(\ref{wpequation})
over all $\xvec$ gives the Waxman and Peck result for this normalization:
\begin{equation}
\Gamma(t+1)= \int \phi(\xvec,t) w_1(\xvec) d^Nx~.
\label{w1bareq}
\end{equation}
We follow Waxman and Peck in using a Gaussian function for the mutation
probability:
\begin{equation}
f(\xvec - \yvec) =
({2\pi m^2})^{-N/2} \exp[-(\xvec-\yvec)^2/2m^2],
\label{fdef}
\end{equation}
where $m^2$ describes the variance of the mutant effects
for a single character.

In addition to this discrete-time model, we will consider
a model in which both mutations and selection occur continuously
in time.
The relation between the discrete and continuous-time models
can be seen explicitly by noting that the differential equation
\begin{equation}
\frac{\partial\phi(\xvec,t)}{\partial t} =
\left [ a(t) - \frac{x^2}{2V} \right ] \phi(\xvec, t)
\end{equation}
has the solution
\begin{equation}
\phi(\xvec, t+\tau) = \exp \left [ \int_{t}^{t+\tau}ds~a(s)
- \frac{x^2}{2V} \tau \right ] \phi(\xvec, t).
\end{equation}
Thus, model Eq.~(\ref{wpequation}) can be written
as a differential equation:
\begin{eqnarray}
& & \frac{\partial\phi(\xvec,t)}{\partial t}  =
~\left [ a(t) - \frac{x^2}{2V} \right ] \phi(\xvec, t)
 \nonumber \\
& &~~~~~~~ +~ \theta \sum_n \delta(t-n\tau) \int d\yvec ~f(\xvec-\yvec)
\left [\phi(\yvec,t) - \phi(\xvec,t) \right ]~,
\label{generalequation}
\end{eqnarray}
where $\tau$ is the interval between mutations.  %%LPK changed
Here, $a(t)$ is determined by the normalization requirement of Eq.~(\ref{norm})
and $\delta(t)$ is the Dirac delta function.
In dynamical systems theory, equations with a sum of
time-delta functions are said to be `kicked'\cite{LL92}.  %%LPK added
When the interval between kicks (here $\tau$) goes to zero, the sum of
delta-function is replaced by its
time average, so that Eq.~(\ref{generalequation}) reduces to:
\begin{equation}
\frac{\partial\phi(\xvec,t)}{\partial t} =
\left [ a(t) - \frac{x^2}{2V} \right ] \phi(\xvec, t)
+ \Theta  \int d\yvec ~f(\xvec-\yvec)
\left [ \phi(\yvec,t)-\phi(\xvec,t) \right ] ~,
\label{leoequation}
\end{equation}
where $\Theta $ is the rate of mutations per unit time,
$\theta/\tau$, %% LPK added
and $a(t)$ is determined by
the normalization requirement, Eq.~(\ref{norm}.)

In the next few sections
we calculate the evolution of $\phi(\xvec, t)$
for different $N$, with an emphasis on the behavior near
$\xvec = 0$ at long times.
In section~\ref{discrete} we examine the discrete-time
model Eq.~(\ref{wpequation}); section~\ref{continuous}
discusses the continuous time equation Eq.~(\ref{leoequation}).
In section~\ref{crossover} we discuss the relationship
between the results for the continuous and discrete-time
equations.

\section{Time Evolution of Discrete Equations}
\label{discrete}
In this section
we consider the discrete-time evolution Eq.~(\ref{wpequation}).
We will solve it
by exploiting
the following observation.
Suppose at some time $t$ $\phi(\xvec, t)$ is a normalized Gaussian
with variance $\alpha$, so that $\phi(\xvec, t) = G_{\alpha}(\xvec)$,
where\footnote{Strictly speaking, $\alpha$ is
the variance of one component of $\xvec$.}
\begin{equation}
G_{\alpha}(\xvec) = (2 \pi \alpha)^{-N/2}
\exp\left[ - \frac{x^2}{2\alpha} \right ] ~.
\label{gaussform}
\end{equation}
Eq. (\ref{wpequation}) then implies that $\phi(\xvec, t+1)$
is the sum of two Gaussians.
Indeed, at any finite time $t$, $\phi(\xvec, t)$
is the sum of finitely many Gaussians.

To see how this works, notice that Eq.~(\ref{wpequation}) involves the
processes of multiplication and
of convolution with a Gaussian.
If one multiplies two Gaussians, one
gets another Gaussian according
to the rule:
\begin{equation}
\label{gaussmul}
G_\alpha(\xvec) G_\beta(\xvec)  =(\frac{\gamma}{2\pi \alpha \beta})^{N/2}
G_\gamma(\xvec)
\mbox{,~~where~~} 1/\gamma= 1/\alpha+ 1/\beta ~.
\end{equation}
Convolution has slightly different rules. A textbook integration yields the
simple result:
\begin{equation}
\int G_\alpha(\xvec-\yvec) G_\beta(\yvec) d^Ny~ =
G_\delta(\xvec)~,
\mbox{~~where~~} \delta= \alpha+ \beta ~.
\label{gausscon}
\end{equation}

Instead of doing integrals at each time step, we can write recursion
relations for the evolution of the
widths and amplitudes of the Gaussians.
We then solve these equations in the limit
$V_s \ll m^2$.
Our method of solution has similarities with that used by
\cite{kingman78} on a simpler model.

To perform the calculation explicitly,
first consider a warm-up problem---the case of no mutations ($\theta=0$),
which is described by the equation:
\begin{equation}
\Gamma(t+1) \phi(\xvec, t+1) = \exp[-x^2/2V_s] \phi(\xvec, t) ~,
\label{nomutateeq}
\end{equation}
where $\Gamma(t+1)$ is defined in Eq.~(\ref{w1bareq}).
We assume that, at time $t$, $\phi(\xvec, t)$ is a normalized Gaussian with
variance $\alpha(t)$:
\begin{equation}
\label{initialgaussian}
\phi(\xvec, t) = G_{\alpha(t)}(\xvec)
\end{equation}
The multiplication process of Eq.~(\ref{gaussmul}) implies that the
value of $\phi(\xvec, t+1)$ is,
once again a Gaussian of the form $G_{\alpha(t+1)}(\xvec)$, but that the
new value of $\alpha$ is given
by
$1/\alpha(t+1) = 1/\alpha(t) + 1/V_s$.
This equation for $\alpha(t)$ has the solution:
\begin{equation}
\alpha(t) = \frac{\alpha_0}{1+\alpha_0 t/V_s} ~,
\label{alpha_t_equation}
\end{equation}
where $\alpha_0 = \alpha(t=0)$.
Thus the population distribution remains Gaussian, and
at long times, its width
scales as $|\xvec| \sim t^{-1/2}$.
As expected, fitness selection continually narrows the distribution.

Now we examine the full Eq.~(\ref{wpequation}) with
$\theta > 0$.
We first calculate
$\phi(\xvec, t=1)$, given that $\phi(\xvec, t=0)$ is a Gaussian.
There are two terms in the equation.
Each involves the multiplication of Gaussians, and the mutation term
also involves a convolution.  By the rules of Eqs.~(\ref{gaussmul}) and
(\ref{gausscon}),
$\phi(\xvec, t=1)$ is the sum of two Gaussians of the form:
\begin{equation}
 \phi(\xvec,1) =
 (1-\theta) G_{\beta_1(\alpha)}(\xvec)  +\theta  G_{\beta_2(\alpha)}(\xvec),
\end{equation}
where the two $\beta$'s are given by:
\begin{equation}
\beta_1(\alpha) = \frac{\alpha}{1+\alpha/V_s}~,
\end{equation}
\begin{equation}
\beta_2(\alpha) = m^2 + \beta_1(\alpha)~,
\end{equation}

It follows immediately that if $\phi(\xvec,t)$
is the sum of finitely many Gaussians,
then $\phi(\xvec, t+1)$ is also the sum of finitely many Gaussians.
Explicitly, we write $\phi(\xvec, t)$ in the form:
\begin{equation}
\phi(\xvec, t) = \sum_i A_i(t) G_{\alpha_i(t)}(\xvec)~,
\label{expandeqn}
\end{equation}
with the normalization condition that
$\sum_i A_i = 1$.
Then $\phi(\xvec, t+1)$ can be written:
\begin{eqnarray}
\Gamma(t+1) \phi(\xvec, t+1) =
(1-\theta) \sum_i B_i(t) G_{\beta_1(\alpha_i(t))}(\xvec)
+ \theta \sum_i B_i(t) G_{\beta_2(\alpha_i(t))}(\xvec) ~,
\label{newphieqn}
\end{eqnarray}
with:
\begin{equation}
B_i(t) = A_i(t)\left [ 1 + \alpha_i(t)/V_s \right ] ^{-N/2}~,
\end{equation}
\begin{equation}
\Gamma(t+1) = \sum_i B_i(t) ~.
\label{endofnewphieqn}
\end{equation}

Eqs.~(\ref{newphieqn}-\ref{endofnewphieqn})
simplify considerably when $V_s/m^2 \ll 1$.
In this case,  the characteristic width of the mutation probability
function,
$f$, is much greater than the width produced  by one step of the fitness
narrowing.  In
this big-mutation-jump limit,
%%LPK add
 $\beta_1(m^2) \approx
V_s$,
$\beta_2(m^2) \approx m^2$, $\beta_1(V_s/n) = V_s/(n+1)$,
$\beta_2(V_s/n) \approx m^2$.\footnote{%
Because the newly-mutated population is
a Gaussian of width determined by
$m^2$ for any distribution,
this approximation is equivalent to the ``house-of-cards''
approximation introduced by \protect{\cite{kingman78}} and
investigated in \cite{turelli84}, \cite{barton89}.
One can show that the results derived in this section are robust
when this approximation is not made, so long as one is the regime
where $V_s \ll m^2$.}
(Here, $n$ is any positive integer.)

Consider the time evolution of an initial distribution consisting
of a Gaussian with variance $\alpha = m^2$.
After one time step, the distribution splits into two
Gaussians, an unmutated population with variance $\beta \approx V_s$
and a mutated population with variance $\beta \approx m^2$.
After two time steps, the population consists of three
Gaussians with $\beta$ values $\approx m^2$, $V_s$, and $V_s/2$.
After $n$ steps, the population consists of $n+1$ Gaussians
with $\beta$ values $m^2$, $V_s$, $V_s/2$, $..., ~V_s/n$.
We define $\alpha_0 = m^2$, $\alpha_i = V_s/i$ for $i \ge 1$.
(Note that these $\alpha$'s are time-independent.)
If we start with an initial distribution which is a
Gaussian of width much greater than $m$, then
after $n$ steps, the distribution consists
of $n+1$ Gaussians with $\beta$ values
$\alpha_0 , ~...,~\alpha_n$.
We describe the time evolution using rate equations
for the populations in these states.
Fig. 1 is a sketch of the transitions that can occur between
the different $\alpha$'s.
Note that at a given time, a state $\alpha_i$
makes transitions to two states, $\alpha_0$ and $\alpha_{i+1}$.

We write $\phi(\xvec, t)$ as:
\begin{equation}
\phi(\xvec, t) = \sum_{i=0}^t A_i(t) G_{\alpha_i}(\xvec) ~,
\end{equation}
where the $A_i(t)$ must obey:.
\begin{equation}
A_0(t>0) = \theta
\label{A_0equation}
\end{equation}
\begin{equation}
A_1(t>0) = \frac{(1-\theta)}{\Gamma(t)}Q A_0(t-1)~,
\label{A_1equation}
\end{equation}
\begin{equation}
A_i(t) = \frac{1-\theta}{\Gamma(t)}
\left [ \frac{i-1}{i} \right ]^{N/2} A_{i-1}(t-1)~~~(i>1,~t>0)~,
\label{A_iequation}
\end{equation}
and
\begin{equation}
\Gamma(t+1) = Q A_0(t) +
\sum_{i=1}^t A_i(t)\left (\frac{i}{i+1}\right )^{N/2} ~.
\label{w1bartequation}
\end{equation}
We have defined $Q = (1+m^2/V_s)^{-N/2}$; note that $Q \ll 1$
in the limit we consider.

First consider the first two steps of the evolution.
Starting with the initial condition
$A_0(t=0) = 1$, $A_i(t=0) = 0$,
one finds:
\begin{eqnarray}
A_0(t=1)=\theta~, \nonumber \\
A_1(t=1)=1-\theta~.
\end{eqnarray}
So far not much has happened.  But a key thing happens at the next step:
\begin{eqnarray}
A_0(t=2)=\theta~,~~~~~~~~~~ \nonumber \\
A_1(t=2)= \frac{\theta Q(1-\theta)}
{\theta Q+(1-\theta)2^{-N/2}}~, \nonumber \\
A_2(t=2)=\frac{(1-\theta)^2 2^{-N/2}}{\theta Q + (1-\theta)2^{-N/2}}~.
\end{eqnarray}
Note that $A_1(t=2) \propto Q \ll 1$.  In fact, for
any $t>2$, all $A_i$'s with $0<i<t$ are proportional to $Q$.
However, as $t \rightarrow \infty$, the number of these
terms diverges.
So it is not obvious whether as
$t \rightarrow \infty$
a solution exists where $A_0(t) = \theta$,
$A_t(t) = O(1)$, and all other $A_i(t)$'s are small,
which would imply
that the Gaussian describing the unmutated population
contains a nonzero fraction of the total population.
We will find that such a solution can exist only when $N>2$.

First we find the steady-state solutions
in the long-time limit $t \rightarrow \infty$.
In steady state, the time arguments on the $A_i$
and on $\Gamma$
can be dropped, and the recursion relations become:
\begin{equation}
A_0 = \theta
\label{ssa0equation}
\end{equation}
\begin{equation}
A_1 = \theta Q v~,
\label{ssa1equation}
\end{equation}
\begin{equation}
A_i = v
\left [ \frac{i-1}{i} \right ]^{N/2} A_{i-1}~~~(i>1)~,
\label{ssaiequation}
\end{equation}
and
\begin{equation}
\frac{1-\theta}{v} =
\theta Q + \lim_{t \rightarrow \infty}
\sum_{i=1}^t A_i\left (\frac{i}{i+1}\right )^{N/2} ~,
\label{ssvequation}
\end{equation}
where we have defined $v = (1-\theta)/\Gamma$.
We explicitly allow for the possibility that
the amplitude
$A_\infty \equiv lim_{t \rightarrow \infty} A_t(t)$ is nonzero.

The solution to Eqs.~(\ref{ssa0equation}-\ref{ssvequation}) is:
\begin{eqnarray}
A_0 = \theta, ~~~~~~~~~~~~~~~ \nonumber \\
A_i = \theta Q i^{-N/2} v^i~,~~~~~~i\ge 1~,
\label{sssolution}
\end{eqnarray}
where $v$ must satisfy:
\begin{equation}
\frac{1-\theta}{v} = A_\infty +
\theta Q \sum_{i=0}^\infty v^i\left (\frac{1}{i+1}\right)^{N/2}~.
\label{ssnormalization}
\end{equation}
Note that $v \le 1$, for otherwise the $A_i$
cannot sum to unity, which is required for normalization
of the probability.\footnote{This inequality is derived in
\protect{\cite{waxman98}}, and is closely related to results
derived by B\"urger and collaborators~
\protect{\cite{burger86}, \cite{burger88}, \cite{burger94},
\cite{burger96}}}
Since as $v$ increases the left hand side of Eq.~(\ref{ssnormalization})
monotonically decreases and the right
hand side monotonically increases,
a solution with $A_\infty = 0$ can exist only if
$\theta Q \sum_{i=0}^\infty \left (\frac{1}{i+1}\right)^{N/2} > 1-\theta$.
Conversely, if this inequality is not satisfied, then
we expect $A_\infty > 0$.  Since
$A_\infty \sim \lim_{t \rightarrow \infty} v^t$,
we require $v=1$ if $A_\infty$ is nonzero.
We show below that such a solution is the long-time limit
of the solution of the time-dependent equations.

First consider the case when $A_\infty \ne 0$.
Since the sum in Eq.~(\ref{ssnormalization})
converges if $N>2$, and diverges otherwise, we
see that $A_\infty \ne 0$ can occur only if $N>2$.
When the sum does converge, the unmutated fitness
peak, whose amplitude is $A_\infty$ and
whose width is given by Eq.~(\ref{alpha_t_equation}),
can contain a nonzero fraction of the total weight.
Evaluating Eq.~(\ref{ssvequation}) with $v=1$, we
find $A_\infty = 1-\theta - \theta Q\zeta(N/2)$, where
$\zeta(N/2)$ is the Riemann zeta function\cite{KCP83,abramowitz72}.

Now we consider the possibility of a solution in which
the unmutated population
constitutes an infinitesimal fraction of the total population
as $t \rightarrow \infty$.
Since $A_\infty=0$, we must have:
\begin{equation}
\frac{1-\theta}{\theta Q} =  \Phi(N/2, v)~,
\label{vequation}
\end{equation}
where $v<1$ and $\Phi(a,x)$ is the polylogarithm function\cite{lewin81}:
\begin{equation}
\Phi(a,x) \equiv \sum_{i=1}^\infty \frac{1}{i^{a}} x^i~.
\label{Phiequation}
\end{equation}

When $a>1$, $\Phi(a, x=1) = \zeta(a)$ is finite, where again
$\zeta(z)$ is the Riemann zeta function.
Therefore, for $N>2$, the
the right hand side of Eq.~(\ref{vequation})
is bounded as $v \rightarrow 1$, and a solution exists only if
$(1-\theta)/(\theta Q) <  \zeta(N/2)$.
In the $Q \ll 1$ limit that we have assumed, this happens only when
$1-\theta$ is very small also.
If this condition is not satisfied, then a nonzero
fraction of the population must be in the unmutated state.

When $N=2$, $\Phi(1, v) = -\log(1-v)$, and
in the limit $\theta Q/(1-\theta) \ll 1$, one finds
\begin{equation}
v \approx 1 - \exp\left [-\frac{1-\theta}{\theta Q} \right ] ~,
\end{equation}
so that:
\begin{equation}
\Gamma \approx (1-\theta)
\left (1 + \exp\left [-\frac{1-\theta}{\theta Q} \right ] \right )~~~~
(N=2).
\end{equation}
When $N=1$, the leading order term as $v \rightarrow 1$
can be calculated by noting that, for all $v < 1$,
$\Phi(\half, v)$ obeys the bounds:\footnote{These bounds follow
because the integrand is monotonically decreasing.  For any
positive integer $j$ and $v \le 1$, one has
$\int_{j}^{j+1} dx~\frac{1}{\sqrt{x}} v^x < \frac{1}{\sqrt{j}} v^j
< \int_{j-1}^{j} dx~\frac{1}{\sqrt{x}} v^x.$}
\begin{equation}
\int_1^\infty dx ~\frac{1}{\sqrt{x}} v^x < \Phi(\half, v)
< \int_0^\infty dx ~\frac{1}{\sqrt{x}} v^x  ~.
\end{equation}
The integrals both converge at their lower limits.
They also converge at their upper limits when $v<1$, but
not when $v=1$.
The behavior near the upper limit can be used to estimate that,
as $v$ approaches $1$,
\begin{eqnarray}
\Phi(\half, v) \sim \int_0^\infty dx ~\frac{1}{\sqrt{x}}
\exp[-x |\ln v |]
\nonumber \\
= \sqrt{\frac{\pi}{|\ln v|}} \int_0^\infty dt~ \exp[-t^2]~
\nonumber \\
= \frac{\pi}{2\sqrt{|\ln v|}}
\nonumber \\
\approx \pi/(2 \sqrt{1-v})~.
\end{eqnarray}
Thus, $v \approx 1 - [(\pi \theta Q)/(2(1-\theta))]^2$, and:
\begin{equation}
\Gamma \approx (1-\theta)
\left (1+\left (\frac{\pi \theta Q}{2(1-\theta)}\right )^2\right )
~~~~(N=1).
\end{equation}

Now we calculate the
function $\phi(\xvec, t\rightarrow \infty)$.\footnote{%
The form of the distribution calculated
by Waxman and Peck (their footnote 29) does
not apply here because they require simultaneous validity
of the inequalities
$\theta V_s/m^2 \ll 1$, $m^2/V_s \ll 1$, and
$N \ll V_s/m^2$.  Our calculation applies
whenever $V_s/m^2 \ll 1$.
}
First we consider the case
when $\phi$ has a $\delta$-function piece.

In this regime, as $t \rightarrow \infty$,
$A_0=\theta$, $A_i=\theta Q(1/i)^{N/2}$
$(i\ge 1)$, and $A_\infty$, the weight in the $\delta-$ function,
is $A_\infty = 1-\theta - \theta Q\zeta(N/2)$,
where $\zeta(N/2)$ is the Riemann zeta function.
Therefore,
\begin{eqnarray}
\phi(\xvec, t \rightarrow \infty)  =
(1-\theta-\theta Q\zeta(N/2))\delta(x)
+ \theta (2\pi m^2)^{-N/2} \exp\left [-\frac{x^2}{2m^2} \right ]
\nonumber \\*
+ \theta Q \sum_{n=1}^\infty (1/n)^{N/2} (2\pi V_s/n)^{-N/2}
\exp\left [ -\frac{n x^2}{2V_s} \right ] ~\nonumber \\*
= (1-\theta-\theta Q\zeta(N/2))\delta(x) +
\theta (2\pi m^2)^{-N/2} \exp\left [-\frac{x^2}{2m^2} \right ]
\nonumber \\*
+  \theta Q (2\pi V_s)^{-N/2} \left (
\exp\left [ \frac{ x^2}{2V_s} \right ] - 1 \right )^{-1}~.~~~~~~~~~~~~~
\end{eqnarray}
Thus we see that when
$\phi(\xvec, t\rightarrow \infty)$ has a $\delta$-function
piece, there is in addition a divergent contribution
at small $x$, proportional to $ 1/x^2$.
When the distribution is smooth,
we have $A_0=\theta$ and $A_i=\theta Q(1/i)^{N/2} v^i$ $(i\ge 1)$,
and $v<1$,
and we obtain:
\begin{eqnarray}
\phi(\xvec, t \rightarrow \infty)  =
\theta (2\pi m^2)^{-N/2} \exp\left [-\frac{x^2}{2m^2} \right ]
\nonumber \\*
+  \theta Q (2\pi V_s)^{-N/2} \left (
v^{-1} \exp\left [ \frac{ x^2}{2V_s} \right ] - 1 \right )^{-1}~.
\end{eqnarray}

The sums that arise here are identical to those that
come up in the calculation of Bose-Einstein condensation
for an ideal Bose gas\cite{degroot50}.
The weight in the $\delta-$function in the genotype distribution,
$A_\infty$, is analogous to the condensate fraction
in the Bose condensation problem.
The parameter $N$, which here describes the number of
traits affected by a mutation, is the
number of spatial dimensions in the Bose gas calculation.
That the superfluid fraction must be zero for a Bose gas in
one and two dimensions is a special case of a
general result in the theory of
phase transitions\cite{mermin66}, \cite{hohenberg67}, \cite{frohlich81}.

We now discuss the time evolution of the population distribution.
We begin with some qualitative remarks.
When $v<1$, the
sum in the solution Eq.~(\ref{ssnormalization})
converges geometrically.
Therefore, in this regime one expects the approach to the
$t \rightarrow \infty$ limit to be exponential.
In the regime where a $\delta-$function contribution is
present as $t \rightarrow \infty$, the $\delta-$function
is the long-time limit of a Gaussian describing the unmutated
population.
The squared width of this Gaussian narrows as $1/t$, so we expect
the approach to the long-time limit to be power-law.
These expectations are supported by the explicit calculation
that we now present.  We, in fact, show that the long-time
corrections to the amplitude of the $\delta-$function
are also proportional to $1/t$.

We again write recursion relations describing the transitions
between the various Gaussians, now allowing for explicit
time dependence in the $A_i(t)$.
Defining $v(t) = (1-\theta)/\Gamma(t+1)$, these recursion
relations are, for $t>0$:
\begin{equation}
\label{recursion1}
A_0(t) = \theta ~,
\end{equation}
\begin{equation}
\label{recursion2}
A_1(t) = Q v(t-1) A_0(t-1) ~,
\end{equation}
\begin{equation}
\label{recursion3}
A_i(t) = v(t-1) \left (\frac{i-1}{i} \right )^{N/2} A_{i-1}(t-1)~,
\end{equation}
\begin{equation}
\label{recursion4}
\frac{1-\theta}{v(t)} = Q A_0(t) + \sum_{i=1}^t A_i(t)
\left ( \frac{i}{i+1} \right )^{N/2}~.
\end{equation}
For the initial conditions $A_0(t=0)=1$ and
$A_{i>0}(t=0)=0$, these recursion relations have the solution:
\begin{equation}
A_i(t) = Q \prod_{\tprime = t-i}^{t-1} v(\tprime)
\left ( \frac{1}{i} \right )^{N/2}A_0(t-i+1)~.
\end{equation}
The self-consistency condition is:
\begin{equation}
\frac{1-\theta}{v(t)} =  \frac{Q}{v(t)} \left [
\sum_{j=1}^{t+1} \left (\frac{1}{j} \right )^{N/2}
A_0(t-j)
\prod_{\tprime=t-j+1}^{t} v(\tprime) \right ] ~.
\end{equation}
Now we separate out explicitly the term $j=t+1$; this is
reasonable because $v(\tprime= 0)$ is larger than all the
other $v$'s by a factor proportional to $1/Q$, yielding:
\begin{equation}
\frac{1-\theta}{v(t)} =
\frac{Q\theta}{v(t)}\sum_{j=1}^t \left ( \frac{1}{j} \right )^{N/2}
\prod_{\tprime=t-j+1}^{t} v(\tprime) +
(1-\theta) \left ( \frac{1}{t+1} \right )^{N/2}
\prod_{\tprime=1}^{t-1} v(\tprime)~.
\label{tdependentsolution}
\end{equation}
Comparison of this result to that of
the steady-state analysis (Eq.~(\ref{sssolution})) reveals
that as $t \rightarrow \infty$
the last term on the rhs is just $A_\infty$,
the amplitude of the $\delta-$function.
Finally, defining $\gamma(t) = \prod_{\tprime = 1}^{t}v(\tprime)$, we find
\begin{equation}
(1-\theta) = Q \sum_{j=1}^t \left (\frac{1}{j} \right )^{N/2}
\frac{\gamma(t)}{\gamma(t-j)} +
(1-\theta) \left ( \frac{1}{t+1} \right )^{N/2} \gamma(t)~.
\end{equation}

Recall that at long times $v(t)$ approaches a limit $v \le 1$.
Therefore, for large $t$, $\gamma(t) = C_\gamma v^t$,
where $C_\gamma$ is a constant,
up to corrections that vanish as $t \rightarrow \infty$, and
\begin{equation}
(1-\theta) = Q \sum_{j=1}^t \left (\frac{1}{j} \right )^{N/2}
v^j + (1-\theta)\left ( \frac{1}{t+1} \right )^{N/2} C_\gamma v^t~.
\end{equation}
When $v$ is strictly less than unity, the convergence is exponential,
as expected from
the qualitative argument given above and in agreement with the
result from the continuous-time equations below.
When $\lim_{t \rightarrow \infty} v(t) = 1$, then, since the second term
on the right hand side is nonzero as $t \rightarrow \infty$, we must
have $\gamma(t) = \prod_{i=1}^t v(t) = C_\delta(t+1)^{N/2}$, where
$C_\delta$ is a constant.  This
implies $v(t) \rightarrow ((t+1)/t)^{N/2} \sim 1+O(1/t)$.
By calculating
the correction to the sum because of this variation in $v(t)$, we
find that the amplitude of the $\delta-$function obeys
$A_t(t)-A_\infty \propto t^{-1}$ as $t \rightarrow \infty$.
When $N=2$, $v(t \rightarrow \infty)$ is exponentially close
to unity, so that the decay of the amplitude in the unmutated
peak up to extremely long times
is governed by the logarithmic divergence of the sum
for $v=1$.  Thus, in this case we expect $A_t(t)$ to decay
logarithmically with $t$.

In figure 2 we plot $A_t(t)$, the fraction of the population
which is unmutated, versus time $t$ for the parameter values
$\theta = 0.2$ and $Q= 0.1$, for $N=1$, $2$, and $3$.
The curves were obtained by numerical iteration of the
recursion relations Eqs.~(\ref{recursion1}--\ref{recursion4}),
starting with the initial condition $A_0(t=0)=1$ and
$A_i(t=0)=0$ for $i \ne 0$.
We find, as expected, that $A_t(t)$ decays to zero for
$N=1$ and $N=2$ but not for $N=3$.
The scales on the plot were chosen to emphasize the
logarithmic decay when $N=2$.

To compute the probability function $\phi(\xvec,t)$ we first calculate the
amplitudes $A_i(t)$, using the
recursion relation, and then sum the corresponding Gaussians
(see Eqs.~(\ref{expandeqn}) and
(\ref{gaussform})) to obtain numerical values of
the probability.  These results are plotted in figures \ref{N1fig},
\ref{N3fig}, and \ref{N3scalefig} for
the parameter-values $\theta = 0.2$ and $Q= 0.1$.
For $N=1$, see figure \ref{N1fig}, we plot
$\phi(\xvec,t)$ against the magnitude of
$x$ for various values of the time, $t$.  The picture shows a
probability distribution which first
gets narrower, but then settles down to a  time-independent behavior for
the largest times. Also
drawn on this plot is the infinite-time probability, derived from
Eqs.~(\ref{sssolution}) and (\ref{ssnormalization}).
The probability $\phi(\xvec,t)$ settles down to this
limiting behavior most slowly near
$\xvec=0$, where fitnesses change slowly with $\xvec$.

For $N=3$, a corresponding calculation shows a probability function which
gets more and more peaked
as time goes on, see figure \ref{N3fig}.   Notice how, for $t=1000$, the
peak sticks up sharply from
a more slowly varying background.  To see how this peaking occurs, we plot
in figure
\ref{N3scalefig} a scaled version of figure \ref{N3fig}.   In this version,
we plot $\phi_s(\xvec_s,t)$
where  $\phi_s(\xvec,t)$ is the result of dividing $\phi(\xvec,t)$ by the
predicted
growth of the peak,  proportional to $t^{N/2}$, giving us a scaled y-variable
$\phi_s(\xvec,t)=\phi(\xvec,t) t^{-N/2}$. The scaled
$x$-variable is
$\xvec_s=\xvec t^{1/2}$.  The resulting plot becomes time-independent for
large times and not-too-large values
of $\xvec_s$.  In this way, we see how the peak continually gets narrower,
and more and more dominates
the small-$x$ behavior.

Now we turn to the question of the role of initial conditions
in the behavior of discrete-time model.
This question is important because
the method of solution of the discrete-time model outlined
in this section
assumes that the initial distribution
$\phi(\xvec, t=0)$ is Gaussian.
Here we present our arguments why the
results of the calculation should not depend sensitively on this
choice of initial condition.

First, note that any initial distribution $\phi(\xvec,t=0)$
which has a width much greater than both $V_s$ and $m^2$
will narrow to two Gaussians of widths
$V_s$ and $m^2$ after one mutation step.
Therefore, the behavior of any broad distribution
will be indistinguishable from that of a Gaussian
of width $\simgt m^2$.

Now we consider more general variation of the initial conditions
by examining the time-dependent Eq.~(\ref{tdependentsolution}).
Every term in the sum arises from a mutation event at some
time $t\ge 0$;
changing the initial conditions changes only the last
term on the right hand side.
In the $t\rightarrow\infty$ limit, the
the amplitude of the $\delta-$function is
obtained by subtracting from unity the fraction of the
population which has mutated at some time.
The long-time limit, $v$, does not
depend on the initial condition; to see this,
note that either $v=1$, in which case $A_\infty$
is obtained by the subtraction just described,
or else $v<1$, in which case $A_\infty=0$ and
the amplitudes of all the terms that depend on the
initial conditions vanish exponentially.
Therefore, the initial conditions do not affect either
the presence or absence of the $\delta-$function or
its steady-state amplitude when it is present.

\section{Continuous-Time Equations}
\label{continuous}
In this section we consider the continuous-time
evolution described by Eq.~(\ref{leoequation}).
In \cite{burger96} it is shown that there is a unique time-independent
solution as $t \rightarrow \infty$ and the possible
solutions are classified using general properties of linear
operators\cite{reed72}.
Our results for the long-time limit
are consistent with theirs.
In addition, our approach enables us to interpret simply
the emergence (or not) of the $\delta-$function peak
as well to address the time-dependence of the approach
to the $t \rightarrow \infty$ limit.

We proceed by Fourier transforming Eq.~(\ref{leoequation}).
We define the quantity
$n(\kvec, t) =
\int d^Nx~ \phi(\xvec, t) \exp[i\kvec\cdot\xvec]$
and obtain:
\begin{equation}
\frac{\partial n(\kvec, t)}{\partial t} =
\left [ a(t) + \frac{1}{2V} {\nabla_{\kvec}^2} \right ]
n(\kvec, t) - \Theta[1-f(\kvec)]n(\kvec,t)~,
\label{ftequation}
\end{equation}
where $f(\kvec)$ is the Fourier transform of Eq.~(\ref{fdef}):
\begin{equation}
f(\kvec) = \exp[-m^2k^2/2]~.
\end{equation}

We must also specify, in addition to the initial
conditions, two boundary conditions.
One boundary condition is determined by the normalization
requirement $\int d^Nx~\phi(\xvec,t)=1$, which yields:
\begin{equation}
n(\kvec=0,t)=1~.
\label{boundarycondition1}
\end{equation}
The second boundary condition is that $n(\kvec,t) \le 1$
for all $\kvec$, which in particular implies that it cannot
diverge as $\kvec \rightarrow \infty$.
This follows from combining the normalization condition
with the non-negativity requirement $\phi(\xvec, t) \ge 0$:
\begin{eqnarray}
n(\kvec,t) =
\int d^Nx~ e^{ i \kvec \cdot \xvec } \phi(\xvec, t)
\le \int d^Nx~ \phi(\xvec, t) = 1~.
\label{boundarycondition2}
\end{eqnarray}

We now solve this equation in
the limit of long times.
Since as $t \rightarrow \infty$ both $n$ and $a$ become
independent of time, we must solve:
\begin{equation}
\left \{ -\frac{1}{2} {\nabla_{\kvec}^2}
- (\Theta V) f(\kvec) \right \} \psi(\kvec) = E \psi(\kvec),
\label{schrodinger1}
\end{equation}
where $\psi(\kvec)\equiv \lim_{t \rightarrow \infty}n(\kvec,t)$
and the eigenvalue $E = V(\lim_{t \rightarrow \infty} a(t)-\Theta$).
Eq.~(\ref{schrodinger1}) is just the time-independent
Schr\"odinger equation\cite{schiff68}.
The parameter $N$, the number of characters affected by
each mutation, here is interpreted as the spatial dimensionality.
As shown in \cite[and references therein]{burger96}
and as discussed here
in the Appendix,
the long-time solution to the population
biology model is given by the lowest energy eigenstate
of this Schr\"odinger equation.

The kinetic energy term $-\frac{1}{2} {\nabla_{\kvec}^2}$
in Eq.~(\ref{schrodinger1}) comes from the fitness selection.
This term, which in $\xvec$-space causes the distribution
to become narrower and narrower, acts
to cause it to become wider and wider in
$\kvec$-space.
The potential energy term $-\Theta V f(\kvec)$ comes from
the mutations.
This potential is constant at large $k$, and has an
attractive piece at small $k$.
If we assume that the potential has a characteristic
scale in $k$, so that $f(\kvec) = F(m\kvec)$ (clearly
the fitness function considered here is of this form), then
we can define $\zvec = m\kvec$ and write the
Schr\"odinger equation in dimensionless form:
\begin{equation}
\left \{ -\frac{1}{2} {\nabla_{\zvec}^2}
- \left ( \frac{\Theta V}{m^2} \right )
F(\zvec) \right \} \psi(\zvec) = \tilde{E} \psi(\zvec),
\label{schrodinger2}
\end{equation}
where $\tilde{E} = E/m^2$.
This equation makes it clear that the nature of the behavior
depends only on the single parameter $(\Theta V/m^2)$.

If only the selection (kinetic energy) term were present,
the solutions to Eq.~(\ref{schrodinger1})
consistent with the normalization condition $\phi(\kvec)=1$
are
\begin{equation}
\psi(\zvec)=\exp[i\pvec\cdot\zvec] ~,
\end{equation}
where $\pvec = \hat{z}\sqrt{2\tilde{E}}$, and $\hat{z}$ is a unit
vector along $\zvec$.
To be consistent with the boundary conditions, we require
$\tilde{E} \ge 0$.
The ground state eigenstate thus has $\tilde{E}=0$, so that
$\psi(\zvec) = 1$.
Fourier transforming this result, we see that
in the absence of mutations, at $t = \infty$ the
entire population has the same genotype.
This result is consistent with our calculation in the
previous section demonstrating that in the absence of mutations
the variance of $\phi(\xvec,t)$ narrows in proportion to $1/$ time.

So now we consider the effects of adding the potential arising
from the mutation term.
This potential is attractive, so that the question
we must address is whether the ground state eigenstate
remains extended
($\psi(\zvec) \rightarrow {\rm constant~  as ~}|\kvec| \rightarrow \infty$)
or whether it results in a bound state with $\tilde{E}<0$, which has
$\psi(\zvec) \rightarrow 0$ exponentially in $k$
as $\kvec \rightarrow \infty$.
In the latter case, the real-space distribution $\phi(\xvec)$
is smooth as $\xvec \rightarrow 0$.
Therefore, if there is a bound state, then only an infinitesimal
fraction of the population is unmutated, whereas if there are
no bound states, then a finite fraction of the population has
a unique genotype.

It has been proven for the Schr\"odinger equation
that any attractive potential, no matter how
small, will lead to a bound state for $N \le 2$\cite{simon76}.
Thus, the population distribution in real space
$\phi(\xvec, t \rightarrow \infty)$ must be smooth
for $N \le 2$.
When $N > 2$ bound states only appear if the potential
is large enough ($\Theta V_s/m^2 > C_{N}$, where
$C_{N}$ is of order unity)\cite{schiff68}.
Therefore, in this case, if the potential is small
(weak mutation effects), then the ground state remains extended:
$\psi(\kvec) \rightarrow constant$ as $\kvec \rightarrow \infty$,
and the real space distribution $\phi(\xvec)$ has a
$\delta-$function piece.
If the potential is large (strong mutation effects),
then there is a bound state, and the distribution $\phi(\xvec)$
is not singular at small $\xvec$.

If all the states are extended, then as
$|\kvec| \rightarrow \infty$ the lowest energy state
obeys ${\nabla_{\kvec}^2} n(\kvec) = 0$ and
has the form $n(\kvec) \rightarrow A + Bk^{-(N-2)}$,
where $A$ and $B$ are coefficients determined by
matching to the solution in the small $|\kvec|$
region.\footnote{This form can be verified for all $N$
by several means, the most straightforward being direct evaluation
using Cartesian coordinates.  A more elegant method is to note that
we are looking for a function $g$ which satisfies
$0 = \nabla^2 g = {\rm div} \cdot {\rm grad}~ g$.
Since ${\rm grad}~ g$ is a current $\jvec$ that is
angle-independent and obeys ${\rm div}\jvec = 0$, then
$\int j~dS_r = 0$, where $S$ is the surface of a sphere of
radius $r$.  Therefore, one must have $j \propto r^{-(N-1)}$,
so that $g \propto r^{-(N-2)}$.}

For almost all potentials both $A$ and $B$ are nonzero;
Fourier transforming the term proportional to $B$ yields:
\begin{eqnarray}
{\rm B~term} \propto \int d\Omega \int_0^\infty
dk~k^{N-1} e^{ikx\cos(\theta)} k^{-(N-2)}
\propto {x}^{-2} ~.
\end{eqnarray}
(Here, $\int d\Omega$ is the integral over angles.)
Thus, just as in our discrete-time solution, we find
associated with a $\delta$-function at $\xvec=0$
a power-law divergence at small $x$, $\phi(\xvec) \propto x^{-2}$.

In the regime where the ground state is bound, there is a finite
energy gap between the ground state and the excited states, which
implies that the the long-time limit is approached exponentially
in time.
When there is no bound state, the energy spectrum of the
Schr\"odinger equation is continuous, which leads to power-law
convergence to the long-time limit.
These points are discussed in greater detail in the Appendix.

\section{Relation between the continuous-time and
the discrete-time models}
\label{crossover}
In this section we discuss the similarities and differences
between the results of our discrete-time analysis valid when
$V_s/m^2 \ll 1$ and those of the continuous-time analysis.
The basis of our discussion is Eq.~(\ref{generalequation}),
which can be used to describe both cases.

Qualitatively, the discrete and continuous models exhibit
very similar behavior.
They both lead to only smooth distributions (a ``normal" phase)
for $N\le 2$, and for $N>2$ display a transition as the
mutation rate is decreased between the
normal phase and a ``condensed" phase, where a finite fraction of
the population has a unique genotype.
In both models, the convergence to the long-time limit
is power law
%% LPK (more specifically, diffusive)
in the condensed
phase and is exponential in the normal phase.
However, there are quantitative differences between the
two models, particularly the detailed dependence on model
parameters.

In the continuous-time model, the form of Eq.~(\ref{schrodinger2})
makes it clear that the behavior depends
only on the single parameter $\Theta V/m^2$.
In contrast, the discrete model that we solve in the limit
$m^2/V_s \gg 1$ (Eqs.~(\ref{A_0equation}-\ref{A_iequation}))
depends on two parameters: $Q = (1+m^2/V_s)^{-N/2}$ and $\theta$.
For $N>2$, the discrete model might or might not have a condensate,
depending on the value of the ratio $\theta Q/(1-\theta)$.
In this regime, the continuous-time model is always
in the condensed phase.

We use Eq.~(\ref{generalequation}) to discuss the
relation between the continuous- and discrete-time models.
One can Fourier transform this equation (as we did in
section \ref{continuous} for the continuous-time model)
and interpret the result as a
quantum-mechanical problem of a particle in a potential
which is applied only at the discrete times
$t=n\tau$ as in Eq.~(\ref{generalequation}).\footnote{%
We take the limit where a very strong potential is applied
for a very short time at each $t_n = n\tau$.}

The discrete model has two parameters
while the continuous model has only one, since
Eq.~(\ref{generalequation})
depends not just on the mean mutation rate $\Theta$
but also on $\tau$, the time between mutation events.
Thus, in principle the discrete-time model has one more
parameter than the continuous-time one.
If $\tau$ is increased by a factor of two, then the
product $\theta V$ ($\theta$ the mutation probability per generation
and $V$ the viability selection parameter) remains constant,
but $V$ itself is halved.
The analysis in section~\ref{discrete},
which is valid when $V/m^2$ is very small, and hence
in the limit when $\tau$ is very long, yields a solution
in which that both $\theta$ and $V/m^2$ enter explicitly
and not just as a product.
Nonetheless,
two models have very similar outcomes.

\section{Comparison with sequence database statistics}
\label{database}
Here we present our investigations of
the statistical properties of DNA sequences
archived in genetic databases.
The aim is to see if observed sequence variability of a
gene is correlated with its degree of pleiotropy.
The focus in this section is on a qualitative prediction
of the model, that genes with a high degree
of pleiotropy should have a narrower distribution of alleles
than those that affect only one trait.
This trend, which is consistent with the mathematical results
of the previous sections, is easy to understand: if a gene of high
fitness is pleiotropic, then because each
mutation affects several characters independently, the
chances are high that a given mutation leads to a large
fitness decrease.

Our first test for possible correlation between
degree of pleiotropy and the probability distribution
describing genetic variation is simply to count
the number of naturally occurring variants
of various genes in the {\it Drosophila}
database FlyBase\cite{flybase98}.
We choose to include naturally occurring alleles,
including spontaneous mutations arising in laboratory
stocks, of the genes analyzed.
Mutagen-induced variants are not included, since there is no
evidence that they exist in natural populations.
Table 1 shows data for nine genes: $brown$, $cinnabar$,
$ecdysone~receptor$, $engrailed$, $fork~head$, $hairless$,
$notch$, $vestigial$, and $white$.
The number of naturally occurring variants is tabulated together
with the length of the primary transcript and the approximate length
of the genomic region encompassing the gene and its flanking
regulatory regions.
Transcript and genomic lengths are taken from the
full-format gene reports in FlyBase.
For genes with multiple transcripts, we report the length
of the longest described transcript.
When the full gene report does not include the genomic length,
this is estimated from the FlyBase molecular map of the gene.
The number of variants per gene is normalized both by
the transcript length and by the genomic length.
In both cases, the values range over approximately
three orders of magnitude.
We estimate the degree of pleiotropy of each of the listed genes
according to the number of tissues/structures and developmental stages
in which the genes are expressed.
This ordering is admittedly arbitrary to some extent, but few
would argue with the conclusion that $fork~head$ (encoding a
transcription factor essential for development of the midgut)
and $engrailed$ (encoding a homeotic gene establishing segmentation)
are more pleiotropic than the eye color mutants $cinnabar$, $white$,
and $brown$.
Inspection of Table 1 reveals a clear inverse correlation
between estimated pleiotropy and number of variants/transcribed length
or variants/genomic length.
This relationship is consistent with the prediction of Waxman
and Peck's model.

A weak point of this approach is that
other factors besides degree of pleiotropy could
lead to the observed differences in the number of variants.
For instance, even two genes which only affect single traits
could have vastly different variabilities, if the traits have
greatly different effects on overall fitness.
For example, a gene that controls eye color may well
have a much smaller overall effect on fitness than
a gene that controls a crucial developmental function.
In the population genetics model, differences of this
type are reflected by different values of
the fitness parameter $V_s$.
Within the model,
differences in overall variability caused by changing $V_s$ can
be distinguished from those caused by changing the degree
of pleiotropy $N$ because they yield very different functional
forms for the distribution function $\phi(\xvec)$.
However, counting mutations yields no information
about the functional form of the distribution function
and thus cannot be used to distinguish between the mechanisms.

Our second strategy for relating degree of pleiotropy
to statistics of archived DNA sequences aims to obtain
information about the form of the probability distribution
for various genes.  It assumes that
two genetic variants whose sequences are highly similar
in sequence space code for genes that are close together
in fitness space.
We use the
BLAST~2 sequence similarity search tool\cite{blast97}
to search the GENBANK and EMBL DNA sequence databases\footnote{These
databases and the BLAST~2 search tool are maintained by
the National Center for Biotechnology Information
(http://www.ncbi.nlm.nih.gov).}
against the 8 gene sequences
listed in Table 2.
These databases contain sequences of genes from many organisms,
though a preponderance of the sequences are from humans.
Given a target sequence, BLAST~2 generates a list of matching
sequences along with scores which measure the degree of similarity.
Because the maximum possible score for a given sequence
is larger for longer sequences, we avoided examining genes
that were either very short or very long.  As can be seen
in table 2, the lengths of the sequences examined
varied by less than a factor of ten.
Only genes were used whose list of matches was truncated
because the score fell below the cutoff value of 40;
no genes were used whose match list was truncated because
the number of matching sequence exceeded
BLAST~2's maximum number of matches of 500.
As shown in table 2, the number of matches obtained
also varies by roughly an order of magnitude
for the genes in the table, and is not obviously correlated
with the length of the gene.

High BLAST~2 scores correspond to
sequences which match the search sequence the most closely.
The scores range from roughly $10^4$ (a sequence matching itself)
to the default cutoff of 40.
We assume that this score is inversely correlated
with the evolutionary distance between the input gene and
the retrieved sequences, so that a gene with
a high fraction of matches with low scores has a broader
probability distribution $\phi(x)$ than a gene with
more matches with high scores.
Thus, if all the matches tend to be
very good, the
gene is considered to be less variable than one with
many poor matches.
We examine the statistics of the closeness of
the matches, in particular the fractions of the total number
of matches that fall into given ranges of scores.

Figure \ref{blastfigure} is a plot of a histogram
of the fraction of the matches in a given score range obtained
versus the inverse of the score for several genes.
We do observe significant variability between these genes;
for example, {\it cytochrome~P-1-450}, which codes for
a general-purpose antioxidant expressed in many tissues
and is thus plausibly highly pleiotropic, has relatively many
very good matches compared to $rhodopsin$, which is quite
specific, coding for a protein essential for vision.
However, we cannot claim that there is an unambiguous
correlation between degree of pleiotropy and these plots.
For example, {\it HOX~A1} has many poor matches, where it might
be expected to be pleiotropic, since it is essential to
many developmental functions.
Thus, though interesting variations in the statistics of DNA
sequences for different genes are found, we have not
been able to demonstrate convincingly a correlation
between degree of pleiotropy and these statistical variations.

The behavior of {\it HOX~A1} can be explained by a process of repeated
duplication and divergence over the course of evolution.
Many of the retrieved genes have assumed distinct but
related developmental functions.  The BLAST~2 analysis presented
here does not account for a shift in score distributions
arising from divergence among members of gene families and superfamilies.
Another fundamental difficulty with our analysis is illustrated by
the case of {\it alpha-tubulin}, which
codes for a protein vital to the formation
of microtubules.
On the one hand, microtubules are expressed in many different
tissues, but on the other hand, all its
functions arise from similar structural properties.
Thus, one could categorize {\it alpha-tubulin} as either highly pleiotropic
(since it is expressed in many tissues) or as non-pleiotropic
(since the function is similar everywhere where it is expressed).
In short, though degree of pleiotropy is defined in the populations
genetic model, it is not clear how to quantify it in terms
of biological function.

There are several additional difficulties and ambiguities
inherent in comparisons between genome sequence data
and population
genetics models of the type considered here.
First, the database scores are based on the quality
of sequence alignments, whereas the population genetics
models define distances in terms of fitness, which is
a phenotypic quantity.
Fitness distance and evolutionary sequence distance may be quite
different; for example, point mutations at different locations
may range in effect from unobservable to lethal.
However, even if there is not a single definite
relationship between sequence distance and fitness distance,
so long as these quantities are positively correlated
with each other, meaningful results may be obtained.
Variability in the relation between the distances
will result in noisy data, but different genes are all
subject to similar uncertainties.
Therefore, it is plausible to hope that, if significant
differences between the sequence variability of two
genes are observed, they reflect
differences in the fitness distribution $\phi(\xvec)$
of the population genetics models.
Another limitation of our analysis is that we consider
the fitnesses of alleles as fixed.
This is not necessarily the case, as many gene products
function as components of multi-protein structures or
pathways\cite{lewontin74}, \cite{dover84}.
This dependency of the fitness value of a specific sequence
variant on the remainder of the genotype is manifested
explicitly by the existence of epistatic interactions
(e.g., \cite{fijneman96}, \cite{vanwezel96}).

In addition, one may worry that the population genetics model
we have used involves a continuously varying phenotype,
whereas sequence variations are discrete.
However, because genes are thousands of base pairs long, there
is still a huge range of variability, and the use of
a continuum model is therefore reasonable.

Finally, as mentioned above,
pleiotropy itself is defined in terms of phenotypic
fitness, and it is not clear how to create an independent
measure that enables one to assign objectively
the degree of pleiotropy $N$ to a given gene.
This is a serious fundamental difficulty that must be overcome
if one is to make meaningful
quantitative statistical analysis of the possible correlations
between the sequence statistics and the degree of pleiotropy.

\section{Discussion}
\label{conclusion}
In this paper we have analyzed some variants of a
population biology model incorporating selection,
mutation, and pleiotropy.  We have focussed on
understanding the circumstances under which at long
times a nonzero fraction of the population has a
unique genotype, and on characterizing the time dependence
of the population distribution in this regime.
We have analyzed both the discrete-time model of
\cite{waxman98} as well as an associated
continuous-time model.  We find:
\begin{enumerate}
\item
In both the discrete and continuous-time models,
a unique genotype can emerge only when $N$, the
number of characters affected by each gene,
is greater than two, a result in agreement with
\cite{waxman98}.
For any $N>2$, the infinite-time population distribution
$\phi(\xvec, t \rightarrow \infty)$
contains a $\delta-$function contribution
when the mutation rate is nonzero but small, but not when
the mutation rate is large enough.
\item
In the regime where
$\phi(\xvec, t \rightarrow \infty)$ has a $\delta$-function
contribution, this $\delta$-function
is accompanied by a $1/x^2$ singularity at small $\xvec$.
\item
The $\delta$-function peak emerges as the limit of
a peak that continually becomes higher and narrower.
Thus, in this regime the convergence to the
$t \rightarrow \infty$ limit is power-law.
\item
In the regime when $\phi(\xvec, t \rightarrow \infty)$
is smooth, the convergence to this distribution
is exponential in time.
\item
The continuous- and discrete-time models exhibit
qualitatively identical behavior, but there are
quantitative differences between them.
\end{enumerate}

Our analysis of the discrete-time equations relies
on the use of Gaussian functions for both the fitness
and mutation terms.  However, our analysis of the
continuous-time model assumes only that the
potential in the quantum mechanics problem is
attractive and short-ranged.
Therefore, we expect our results to apply qualitatively for
a large class of different mutation terms.

Other mechanisms giving rise to genetic
diversity such as antagonistic pleiotropy,
genetic drift, and
discreteness of alleles change qualitatively the
nature of the equations describing the system.
It will be interesting to see whether
the ``condensation"
phenomenon investigated here is robust when these
effects are taken into account.

In addition to our mathematical analysis, we
also attempted to assess whether the degree of pleiotropy
of genes resulted in systematic trends in the degree of variation
they exhibited in the mutations documented in online genetic databases.
These attempts to correlate the degree of pleiotropy with
the statistics of the sequence variations were inconclusive.

Available database resources do not allow unbiased sampling
of the sequence variation present in a natural population.
Such a database is likely to emerge as the Human
Genome Project's present initiative to sample the
extent of sequence variation in a wide array of genes in the
American population progresses.

\section{Acknowledgments}
We are grateful for support by the
National Science Foundation, Grant DMR 96-26119 (SNC),
and the Office of Naval Research,
Grant N00014-96-1-0127 (LPK).
SNC and RDB thank J.J. Sohn for useful suggestions.

\vskip .5cm

\section*{Appendix: The relation between eigenstates of the
Schr\"odinger
equation and the long-time behavior of the continuous-time
population genetics model}
In section~\ref{continuous} above we use the
result that the long-time behavior of Eq.~(\ref{ftequation})
is given by the lowest energy eigenstate of the associated
Schr\"odinger Eq.~(\ref{schrodinger1}).
This result has been shown by B\"urger and
collaborators\cite[and references therein]{burger96};
for completeness we derive it here and emphasize
the implications for the approach to
the long-time behavior in the population biology model.

We assume that we know the solution of the
time-independent problem, which is a set of eigenstates
$u_n(\xvec)$ which satisfy:
\begin{equation}
 - \half{\nabla_z}^2  u_n(\zvec) +
V(\zvec)u_n(\zvec) = E_n u_n(\zvec)~.
\end{equation}
The eigenstates $u_n(\zvec)$ are complete (any function
can be written as a linear combination of them)
and can be chosen to be orthonormal:
\begin{equation}
\int d^Nz~u_{n_1}^*(\zvec)u_{n_2}(\zvec) = \delta_{n_1,n_2}~,
\label{orthonormality}
\end{equation}
where the star denotes complex conjugate.
This normalization is usual in quantum mechanics\cite{schiff68}.

We need to examine the time-dependent
Eq.~(\ref{ftequation}); using scaled variables, it is written as:
\begin{equation}
-\frac{\partial \psi(\zvec, t)}{\partial t} =
- \left [ a(t) + \frac{1}{2} {\nabla_{z}^2} \right ]
\psi(\zvec, t) + V(\zvec)\psi(\zvec,t)~,
\end{equation}
where
\begin{equation}
a(t) = \left [ -\half \nabla_{z}^2 + V(z) \right ]
\psi(\zvec, t) |_{\zvec=0}~.
\end{equation}
We write the wave function at time $t_0$, $\psi (\zvec, t_0)$,
as a linear combination of eigenstates:
\begin{equation}
\psi(\zvec, t_0) = \sum_i b_i(t_0) u_i(z)~.
\label{psiform}
\end{equation}
The condition $\psi(\zvec=0, t_0)=1$ is enforced
by requiring $\sum_i b_i(t_0)u_i(\zvec=0) = 1$.
By substituting the form (\ref{psiform}) for $\psi$
and using the orthonormality of the eigenstates, one finds
equations for the time-evolution of the coefficients $b_i$:
\begin{equation}
-\frac{db_i(t)}{dt} = b_i(t) \left [  E_i - a(t) \right ].
\end{equation}
Integrating this equation yields:
\begin{equation}
b_i(t) = b_i(t_0) {\mathcal A}(t) e^{-E_it}~,
\end{equation}
where ${\mathcal A}(t) = \exp \left [\int_{t_0}^t ds~a(s) \right ]$.
Thus the ratio of the weights of any two eigenstates $i_1$
and $i_2$ is given by:
\begin{equation}
\frac{b_{i_1}(t)}{b_{i_2}(t)} = \frac{b_{i_1}(t_0)}{b_{i_2}(t_0)}
\exp\left [-(E_{i_1}-E_{i_2})t \right ]~.
\label{bresult}
\end{equation}
At long times the exponential factor ensures that the
state with the largest weight is the
one with the lowest energy.

If there is a nonzero energy gap between the lowest and
second-lowest energy eigenstates (no $\delta$-function
in $\phi(\xvec)$), then Eq.~(\ref{bresult}) implies that
the approach to the
long-time limit is exponential in time.
If there is no bound state, then all the states are extended
and the spectrum is continuous;
the $E_i$ take the form $E_i = q_i^2/2$, where $q_i$
has the dimension of a wavevector in $\zvec$ space,\cite{schiff68}
and hence a distance in the original $\xvec$ space.
Therefore,
\begin{equation}
\psi(\zvec, t) = {\mathcal A}(t) \sum_i b_i(t_0) e^{-q_i^2t/2}~.
\end{equation}
The relative weight of the eigenstate $i$ remains large
until $q_i \sim \sqrt{1/t}$.
Thus, the $\delta$-function at infinite times emerges
from a peak that is narrowing, having
a width proportional to $\sqrt{1/t}$.

%%%%%%%%%%%%%%%%%%%%%%%%%%%%%%%%%%%%%%%%%%%%%%%%%%%%%%%%%%%%%%%%%%%%%%%
\newpage

\begin{table}
%{\bfseries Table 1:  {\it Drosophila} mutations}\\[1ex]
\label{flytable}
%\begin{center}
\begin{tabular}{|c|c|r|c|c|c|r|} \hline
~           &           &     &     &   &    &            \\
\rb{~~gene} &  \rb{PR}  & \rb{$N_V$}  & \rb{$L_T$}
            &  \rb{$L_G$}  &  \rb{$N_V/L_T$}  &  \rb{$N_V/L_G$}~~~~  \\
\hline
$fork~head$    & 1     & 1     & 4.0  & 13 - 60  
               & $2.5 \times 10^{-4}$ & $(1.7 - 7.7) \times 10^{-5}$   \\
$engrailed$    & 2     & 19    & 3.6   & 200       
               & $5.3 \times 10^{-4}$  & $9.5 \times 10^{-5}$ \\

$ecdysone~
receptor$      & 3     & 13    & 5.5  & 35
               & $2.4 \times 10^{-3}$  & $3.7 \times 10^{-4}$    \\

$notch$	       & 4     & 70    & 15   & 80 - 150
               & $4.7 \times 10^{-3}$ & $(4.7 - 8.8) \times 10^{-4}$ \\

$vestigial$    & 5     & 55    & 3.8  & 46
               & 0.014                & $1.2 \times 10^{-3}$ \\

$hairless$     & 6     & 17    & 6.0  & 8
               & $2.0 \times 10^{-3}$ & $2.1 \times 10^{-3}$ \\

$cinnabar$     & 7     & 22    & 2.5   & 15        
               & $8.8 \times 10^{-3}$  & $1.5 \times 10^{-3}$  \\

$white$	       & 8     & 247   & 6.0  & 48 - 200
               & 0.041                & $(1.2 - 5.1) \times 10^{-3}$ \\

$brown$	       & 9     & 61    & 3.0  & 140
               & 0.020                & $4.4 \times 10^{-4}$ \\

\hline
\end{tabular}
\vspace {1cm}
\caption{Table of number of naturally-occuring mutations for
different {\it Drosophila} genes.  PR is pleiotropy rank, estimated
using the number of tissues/structures and developmental stages in
which the genes are expressed.  $N_V$ is the number of
naturally-occurring variants of the gene, $L_T$ is the transcript
length, and $L_G$ is the genomic length.  All lengths are
given in kilobases.
}
\end{table}

\begin{table}
%{\bfseries Table 2:  Genes examined using BLAST~2}\\[1ex]
\label{blasttable}
\begin{tabular}{|l|r|r|} \hline
gene        &  length  & number of matches \\
\hline
{\it cytochrome~P-1-450}	  & 2565         & 81           \\
$dystrophin$      		  & 13957        & 216          \\
$epidermal~growth~factor~receptor$  & 2660         & 57 		\\
{\it HOX A1} (human)    	  & 2595         & 380          \\
$huntingtin$	                  & 10348        & 108          \\
$iodothyronine~deiodinase$	  & 2222         & 159          \\
$rhodopsin$	                  & 6953         & 208          \\
{\it alpha-tubulin}	          & 1596         & 488           \\
\hline
\end{tabular}
\vspace {1cm}
\caption{Genes examined using BLAST~2 sequence similarity search
tool.  Lengths of genes are given in units of the number of bases.}
\vskip 1cm
\end{table}

\suppressfloats

% the figures follow here
\newpage
\begin{figure}
\epsfxsize = \hsize
\centerline{\epsfbox{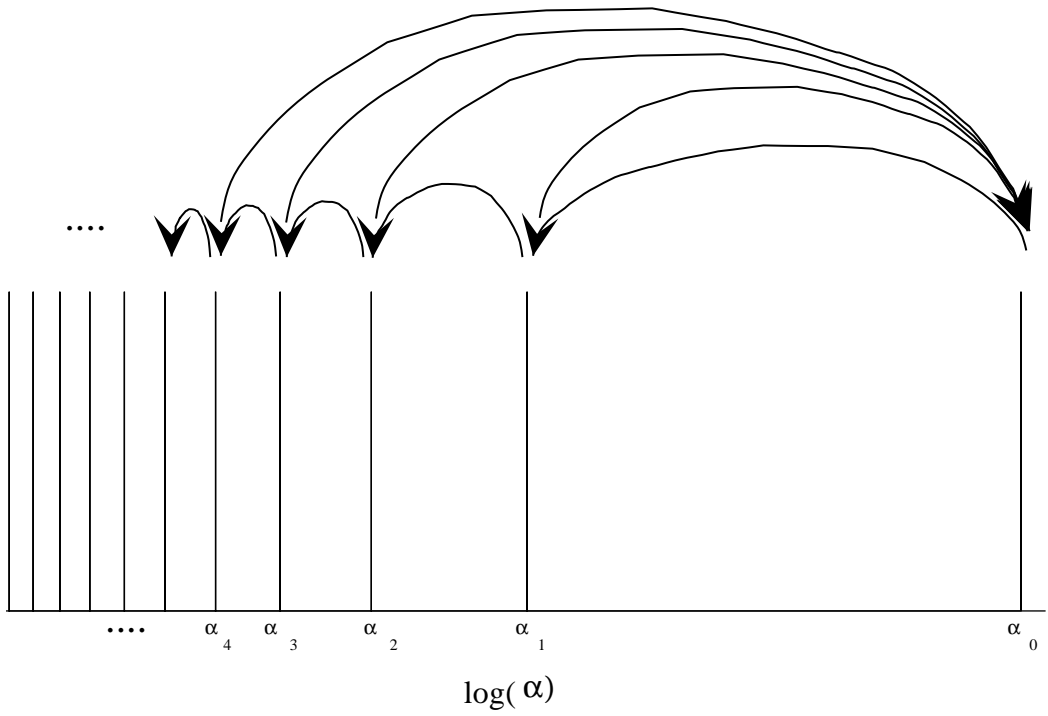}}
\vspace {1cm}
\caption{Sketch of transitions between $\alpha_i$'s in the limit
$V_s/m^2 \ll 1$.  Each $\alpha_i$ corresponds to a Gaussian in
$\phi(\xvec,t)$.}
\end{figure}

\begin{figure}
\epsfxsize = \hsize
%\centerline{\epsfbox{fig2_10000d.eps}}
\centerline{\epsfbox{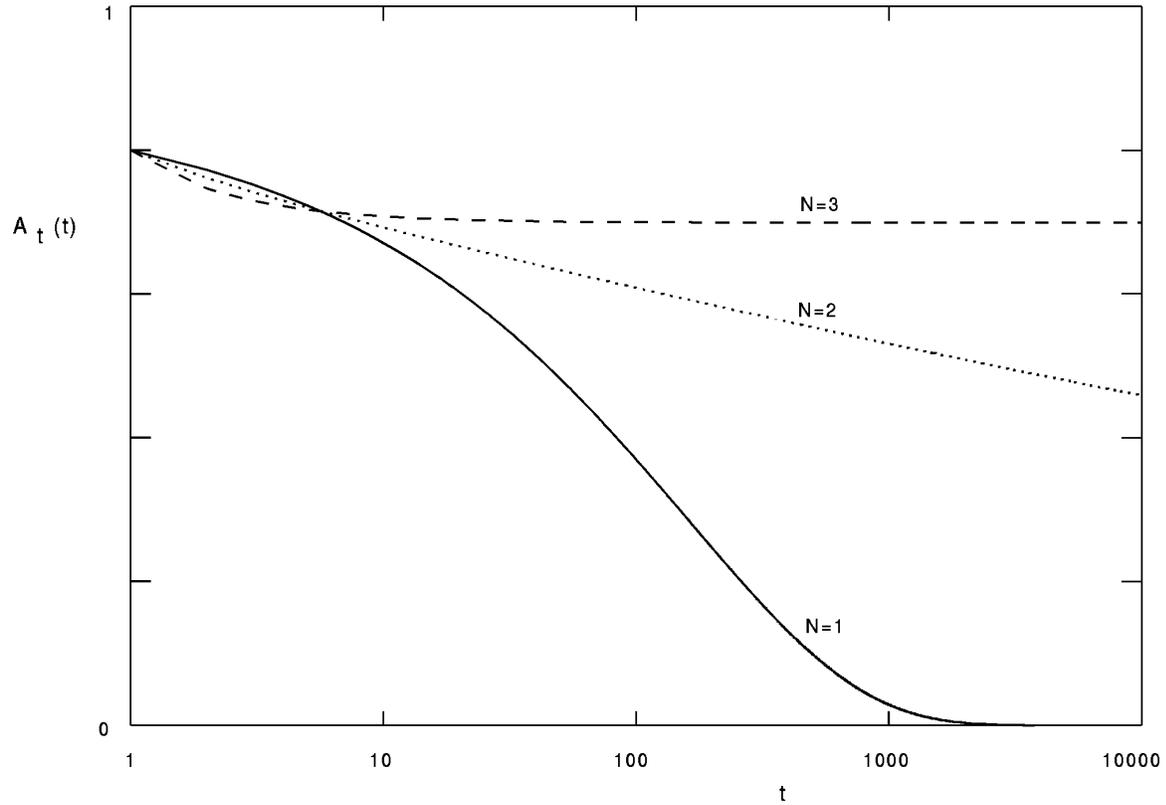}}
\vspace {1cm}
\caption{Plot of the fraction of the population which is
unmutated, $A_t(t)$, versus time $t$ for the discrete-time
model with parameter values $Q=0.1$, $\theta=0.2$,
obtained by numerical iteration of
Eqs.~(\ref{recursion1}--\ref{recursion4}) starting
with $A_0(t=0)=1$, $A_{i\ne 0}(t=0)=0$.
For $N=1$ and $2$, $A_t(t)$ decays so that
$A_\infty \equiv \lim_{t \rightarrow \infty} A_t(t) = 0$,
while for $N=3$, $A_\infty$ is nonzero.
}
\end{figure}

\begin{figure}
\epsfxsize = .9\hsize
%\centerline{\epsfbox{G1rotateda.eps} }
\centerline{\epsfbox{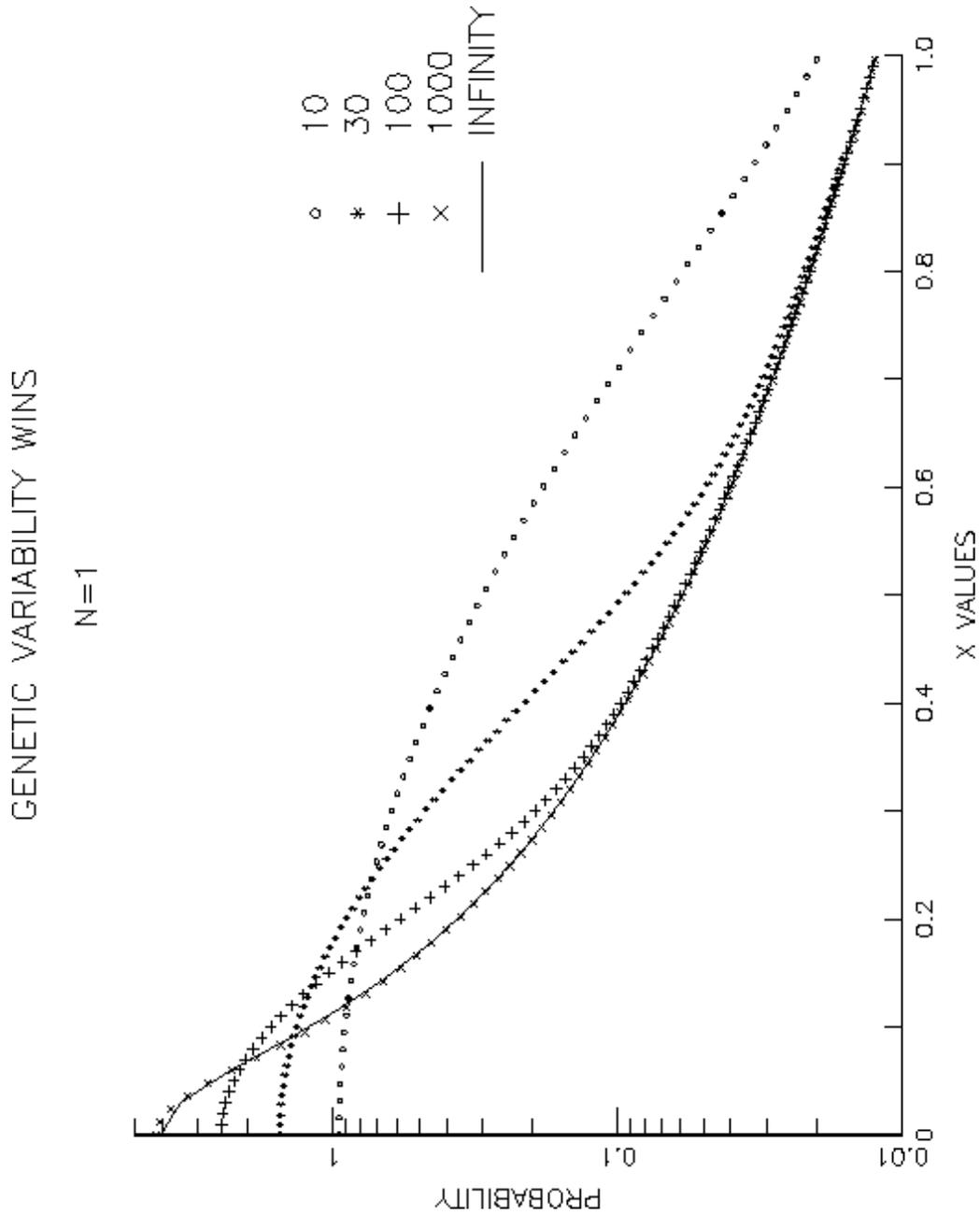} }
\vspace{1cm}
\caption{Plot of probability function $\phi(\xvec,t)$ against the magnitude
of $\xvec$ for $N=1$. The
numerical results for times 10, 30, 100, and 1000 are calculated by summing
Gaussians with
weights computed from the recursion relations. This set of curves is
compared with the expected
infinite-time result, calculated as a solution to
Eqs.~(\ref{sssolution}) and (\ref{ssnormalization}).
For large times, the two types of calculations agree quite well.  }
\label{N1fig}
\end{figure}

\begin{figure}
\epsfxsize = .9\hsize
%\centerline{\epsfbox{G3rotateda.eps}}
\centerline{\epsfbox{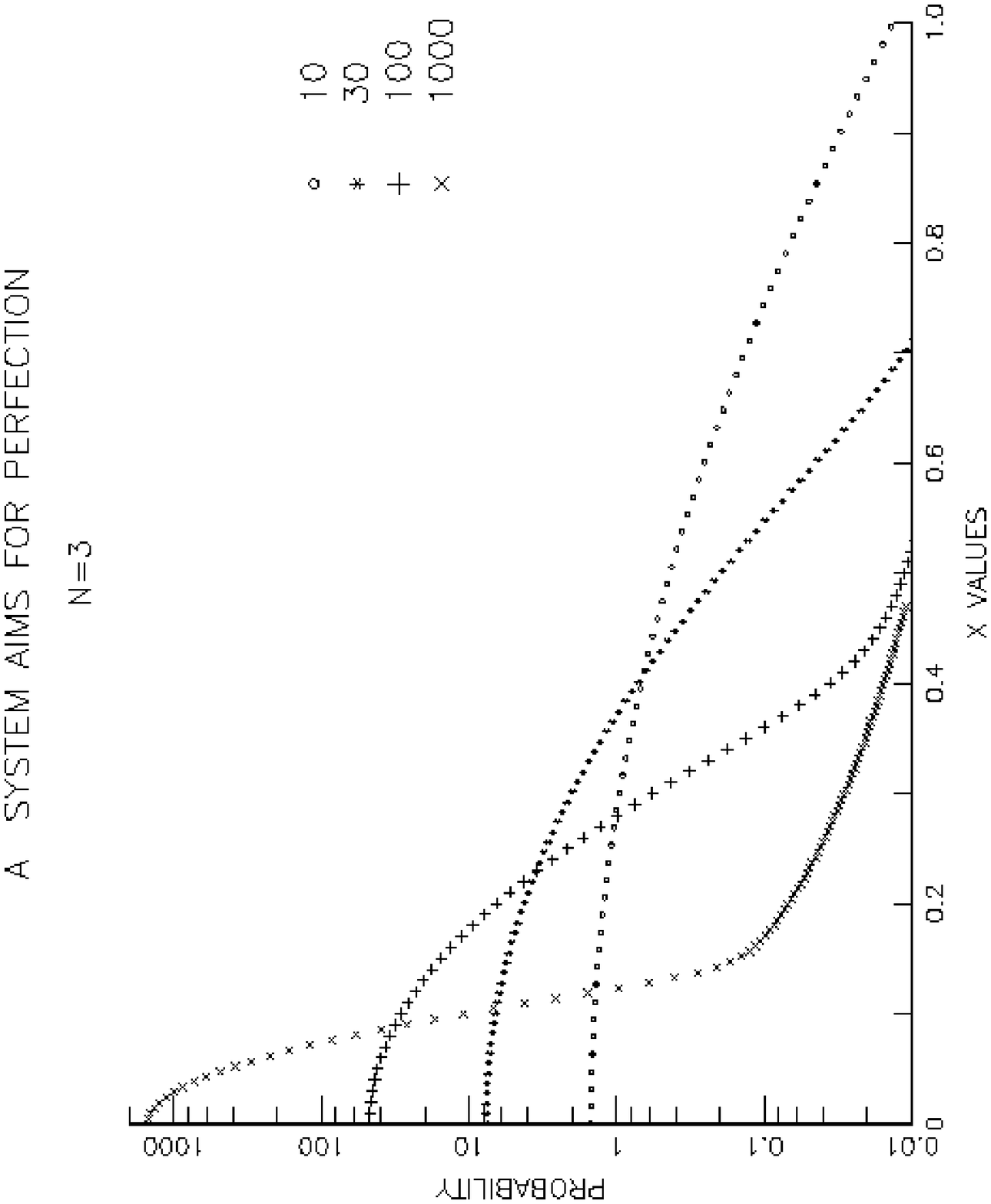}}
\vspace {1cm}
\caption{Plot of probability function $\phi(\xvec,t)$ against the magnitude
of $\xvec$ for $N=3$. The
numerical results are shown for times 10, 30, 100, and 1000.  In contrast
to the $N=1$ case, these curves
contain a central peak which continues to narrow and to grow for large t.   }
\label{N3fig}
\end{figure}

\begin{figure}
\epsfxsize = .9\hsize
%\centerline{\epsfbox{G3srotateda.eps}}
\centerline{\epsfbox{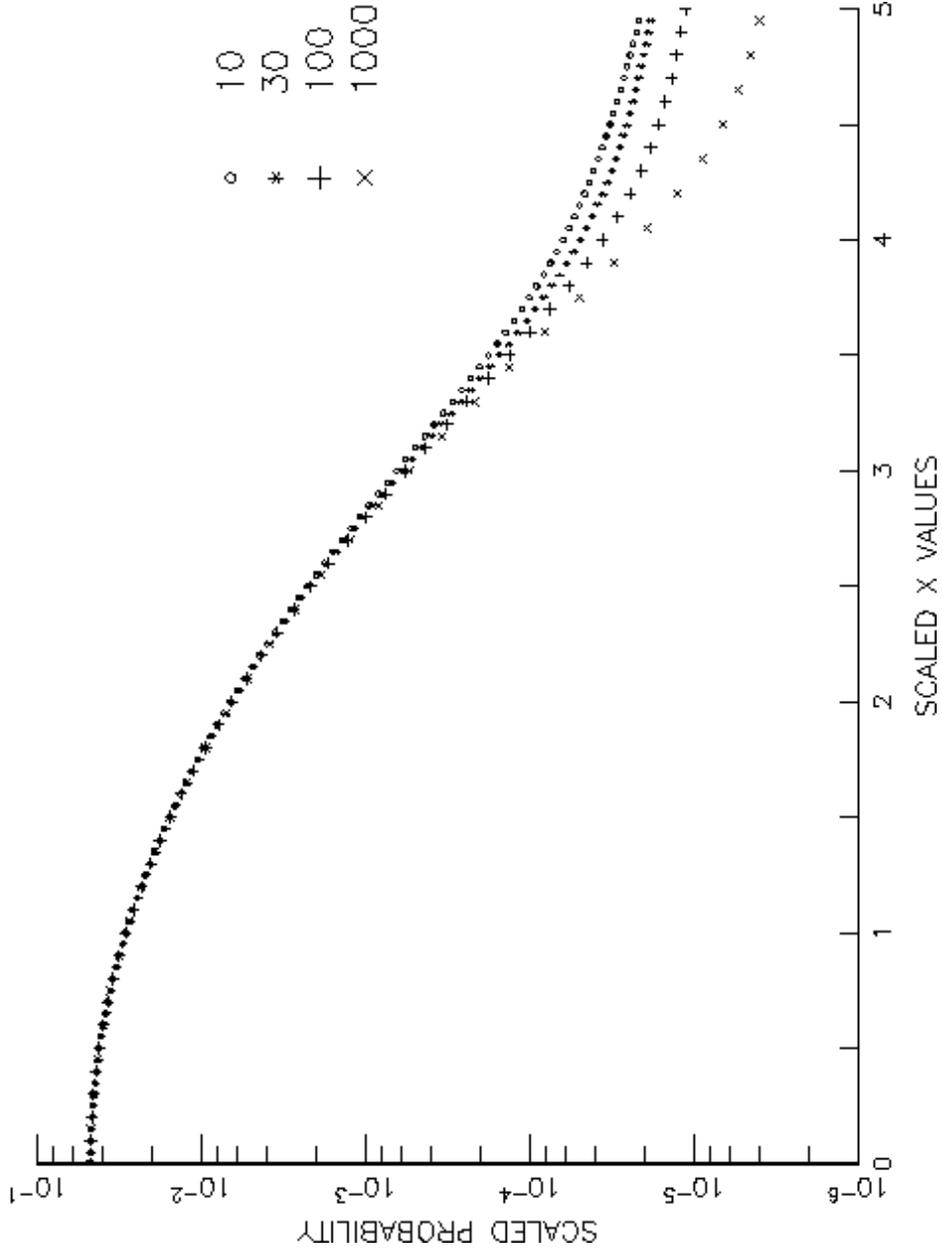}}
\vspace {1cm}
\caption{Scaled plot of probability function versus the magnitude of $x$
for $N=3$.  The scaling is
picked to make the central peak show a time independent behavior
in the new coordinates. The
abscissa is $x t^{1/2} $, so that for larger time the picture focuses upon
smaller values of $x$.  The
ordinate is
$\phi(\xvec,t) t^{-N/2}$, and thereby the picture focuses upon increasingly
concentrated distributions
for larger t.
Numerical results are shown for times 10, 30, 100, and 1000.  In these
coordinates, the figure shows (as
expected) an approach to a constant value at large times.   }
\label{N3scalefig}
\end{figure}

\begin{figure}
\centerline{\epsfbox{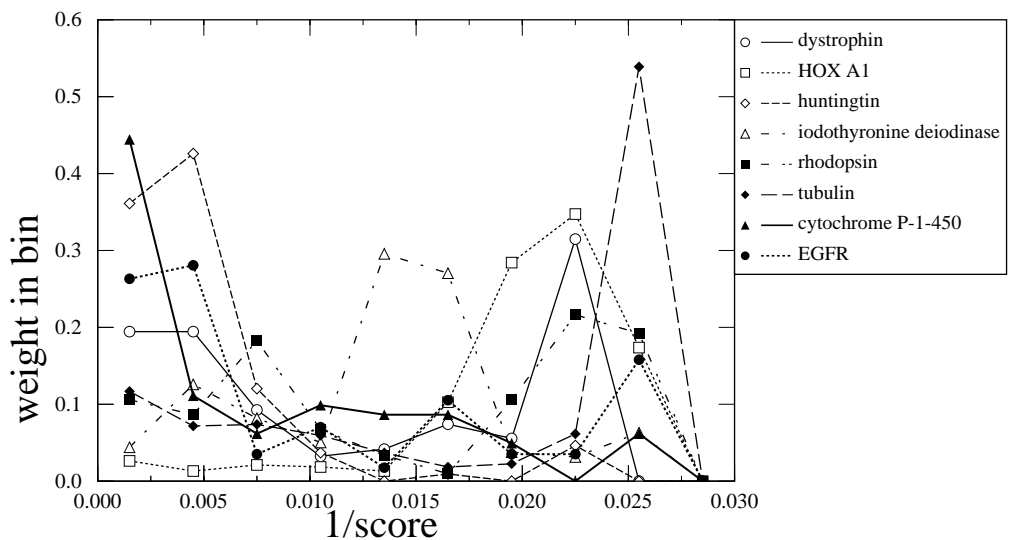}}
\epsfxsize = 1.1\hsize
%\epsfysize = 1.5\vsize
%\centerline{\epsfbox{compressed_figures/blastgraph1.new.ps}}
\vspace {1cm}
\caption{Histogram plot of the fraction of matches
with inverse scores in a given range, as a function
of inverse score, for several genes with differing
degrees of pleiotropy.
Significant differences in the statistics of these
matches are observed; for example, {\it cytochrome~P-1-450}
has a very large percentage of matches with high scores,
whereas the matches for $rhodopsin$ tend to have low scores.
}
\label{blastfigure}
\end{figure}

\end{document}